\documentclass[aps,prb,twocolumn,showpacs,preprintnumbers,superscriptaddress,floatfix]{revtex4-1}
\usepackage{graphicx}
\usepackage{amssymb}
\usepackage{amsfonts}
\usepackage{amsmath}
\usepackage{color}
\definecolor{MyDarkBlue}{rgb}{0,0.08,0.45}
\definecolor{pinegreen}{rgb}{0.0, 0.47, 0.44}
\definecolor{orange}{rgb}{1.0, 0.65, 0.0}

\newcommand{\be}{\begin{equation}}
\newcommand{\ee}{\end{equation}}
\newcommand{\bea}{\begin{eqnarray}}
\newcommand{\eea}{\end{eqnarray}}
\newcommand{\baa}{\begin{align}}
\newcommand{\eaa}{\end{align}}

\newcommand{\ket}[1]{|#1\,\rangle}

\begin{document}
\preprint{APS/123-QED}
\title{Koopmans-compliant spectral functionals for extended systems}

\author{Ngoc Linh Nguyen} \email{linh.nguyen@epfl.ch} 
\affiliation{Theory and
Simulations of Materials (THEOS), and National Centre for Computational Design
and Discovery of Novel Materials (MARVEL), \'Ecole Polytechnique F\'ed\'erale
de Lausanne, 1015 Lausanne, Switzerland} 
\author{Nicola Colonna}
\affiliation{Theory and Simulations of Materials (THEOS), and National Centre
for Computational Design and Discovery of Novel Materials (MARVEL), \'Ecole
Polytechnique F\'ed\'erale de Lausanne, 1015 Lausanne, Switzerland}
\author{Andrea Ferretti} 
\affiliation{Centro S3, CNR--Istituto Nanoscienze, 41125 Modena, Italy} 
\author{Nicola \surname{Marzari}}
\affiliation{Theory and Simulations of Materials (THEOS), and National Centre
for Computational Design and Discovery of Novel Materials (MARVEL), \'Ecole
Polytechnique F\'ed\'erale de Lausanne, 1015 Lausanne, Switzerland}
\date{\today}

\begin{abstract}

Koopmans-compliant functionals have been shown to provide accurate spectral properties for molecular systems; this accuracy is driven by
the generalized linearization condition imposed on each charged excitation - i.e. on changing the occupation of any orbital in the system,
while accounting for screening and relaxation from all other electrons.
In this work we discuss the theoretical formulation and the practical implementation of this formalism to the case of extended systems,
where a third condition, the localization of Koopmans' orbitals, proves crucial to reach seamlessly the thermodynamic limit.
We illustrate the formalism by first studying one-dimensional molecular systems of increasing length. Then, we consider the
band gaps of 30 paradigmatic solid-state test cases, for which accurate experimental and computational results are available.
The results are found to be comparable with the state-of-the-art in diagrammatic techniques (self-consistent many-body perturbation theory with vertex corrections), notably using just a functional formulation for spectral properties
and the physics of the generalized-gradient approximation;
when ionization potentials are compared, the results are roughly twice as accurate.
\end{abstract}

\maketitle

\section{Introduction}
%
Accurate first-principles predictions of spectral properties --- such as band gaps or photoemission spectra --- attract considerable 
attention because of their critical impact on the design and characterization of optical and electronic devices, especially for 
solar energy harvesting and conversion.~\cite{pham_modelling_2017}
To date, the most common approaches to compute these quantities in extended systems are based on many-body perturbation theory (MBPT) using Green's function based approaches (such as the GW approximation)~\cite{Onida2002,Hedin1965} or wavefunction-based methods like coupled cluster~\cite{RevModPhys.79.291} or quantum Monte Carlo~\cite{RevModPhys.73.33} with GW being considered for the case of solids a good compromise between accuracy and computational costs. 
Nevertheless, these high-level methods are still significantly limited in system size and complexity, due to their computational costs, and sometimes even their
accuracy; for these reasons, simpler methods based on Kohn-Sham density-functional theory (KS-DFT), possibly including some fraction of non-local exchange, are still frequently employed to evaluate approximately the spectral properties of nanostructures, interfaces, or solids.

Formally, exact KS-DFT delivers exactly the highest occupied molecular orbital (HOMO) level, since the latter determines the long-range 
decay of the charge density, that needs to be described correctly~\cite{levy_exact_1984,almbladh_exact_1985} (see also Ref.~\onlinecite{PhysRevB.56.16021} and references therein for an in-depth discussion);
all other spectral properties are outside the domain of KS-DFT. The lowest unoccupied molecular orbital (LUMO) in particular is not meant to be correctly positioned, leading to incorrect HOMO-LUMO gaps, although when it becomes infinitesimally occupied it needs to morph into the correct HOMO, explaining why the exact KS potential has
a derivative discontinuity as a function of orbital occupations.~\cite{Perdew-Levy_1983,sham_density-functional_1983}
When dealing with approximate exchange-correlation energy functionals, KS-DFT (especially in the local or generalized gradient approximations) usually underestimates the first ionization potential (IP) and overestimates the first electron affinity (EA), as determined by the negative of the frontier orbital energies. These considerations extend also to solids, and e.g. the band gap energy E$_{\rm g}$ is often greatly underestimated. 

Such failures have been connected to the deviation from piecewise linearity (PWL) of the total energy functional as function of particle number, and the associated lack of derivative discontinuity at integer numbers. First, the deviation from PWL has been suggested~\cite{Cococcioni2005, kuli+06prl,mori-sanchez_many-electron_2006, cohen_insights_2008,Mori-Sanchez_2008} as a definition of electronic self-interaction errors (SIEs)~\cite{perdew_self-interaction_1981}, and in recently developed functionals, such as range-separated~\cite{stein_fundamental_2010,refaely-abramson_quasiparticle_2012} or dielectric-dependent hybrid functionals~\cite{PhysRevB.89.195112, PhysRevX.6.041002}, PWL has been recognized as a critical feature to address. 
The criterion of piecewise linearity was in particular chosen as a key feature by some of us to introduce the class of Koopmans-compliant (KC) functionals~\cite{Dabo2009,Dabo2010,psik_koopmans, Dabo2013, Borghi_PRB_2014}, that enforce a generalized criterion of PWL extending
to the entire manifold the linearizarion criterion of the DFT+Hubbard $U$ approach~\cite{Cococcioni2005, kuli+06prl}. The accuracy in reproducing
spectral properties~\cite{Dabo2013,Borghi_PRB_2014,nguyen_first-principles_2015,Nguyen_JCTC_2016}, in addition to a potential energy surface that preserves exactly or slightly improves~\cite{Nguyen_JCTC_2016} the base functional (typically, using the PBE~\cite{perdew_generalized_1996} approximation), highlight the role of KC functionals as approximations to the exact spectral functional, able to reproduce spectral properties in addition to total energies.~\cite{ferr+14prb}

We briefly summarize here the framework for the KC class of spectral functionals~\cite{Dabo2009, Dabo2010}, to set the discussion; a detailed
description can be found in Refs.~\onlinecite{Borghi_PRB_2014,Borghi_PRB_2015}. The three core concepts that underpin their formulation are those of {\it linearization, screening,} and {\it localization}.

For the first core concept of {\it linearization}, we note that
in a canonical representation (i.e. of orbitals that diagonalize the Hamiltonian) the condition of PWL when extracting an electron from an orbital is naturally akin to that of removing - heuristically - self-interaction from the functional, while also imposing a correct description of charged excitations. The removal of self-interaction can be understood by the fact that KC functionals introduce and impose the condition
that the expectation value of an orbital to be independent from its own occupation - this is our chosen definition of being self-interaction free~\cite{Dabo2010}. One should note that when an orbital is not the HOMO, removing a fraction of an electron from a deep level requires one to constrain that
corresponding orbital to be frozen, and all others to remain orthonormalized to it as the electron is removed (this is the procedure followed
when constructing KC functionals~\cite{Borghi_PRB_2014}). This requirement, in a
canonical representation and through Janak's theorem, is equivalent to a PWL condition for any chosen orbital.
The correct description of charged excitations can be understood by nothing that PWL, in combination as above with Janak's theorem and the frozen-plus-orthonormalization 
constraint, enforces the expectation value of the Hamiltonian to be equal to 
the energy difference between the system with N electrons and that with N-1 electrons, where one electron 
has been removed from the orbital. Such functionals thus aim to describe excited state properties.
In practice, Janak's theorem is never invoked, and the KC formulation postulates what should be the form of the self-interaction free functional (see Eqs.~(4) and~(5) in Ref.~\onlinecite{nguyen_first-principles_2015} and Eq.~(4) in Ref.~\onlinecite{Borghi_PRB_2014}) under the generalized self-interaction constraint
discussed above. The outcome is an orbital-dependent (actually, an orbital-density-dependent) class of functionals that,
akin to the Perdew-Zunger self-interaction functional, are minimized by a set of ``variational'' orbitals, to be contrasted with the
``canonical'' orbitals that diagonalize the $\Lambda$ matrix of orthonormality constraints  (see e.g., Refs.~\onlinecite{Borghi_PRB_2014, nguyen_first-principles_2015, vydrov_tests_2007, pederson-jcp-1984, PhysRevB.77.155106,lehtola_variational_2014}). The variational orbitals minimize the functional, while the canonical orbitals provide the spectra of charged
excitations.

The second core concept of {\it screening} arises from the need to describe and account for the response of the rest of the electronic system as the occupation of one orbital is changed; in order to impose the generalized condition of Koopmans' compliance (i.e. of PWL, under the conditions described above) one needs to take
into account the relaxation taking place in all other orbitals as an electron is removed. In the first applications~\cite{Dabo2010,Dabo2013,Borghi_PRB_2014,nguyen_first-principles_2015} this screening was approximately accounted for using
one single screening coefficient for all filled states, and one for all empty states, being determined respectively by the condition that
the HOMO of the neutral system be equal to the LUMO of the cation (screening coefficient for all filled states), and that the
LUMO of the neutral system be equal to the HOMO of the anion (screening coefficient for the empty states). While this approximation can
be satisfactory and even accurate for small, simple molecules, the formalism calls for orbital-dependent screening. These screening coefficients can be calculated
using finite differences (as done here, and detailed below), or more elegantly using linear-response theory (detailed in Ref.~\onlinecite{Nicola_unpublished}).

The third concept of {\it localization} becomes truly determinant in the thermodynamic limit, i.e. for extended systems: the condition of Koopmans'
compliance not only works better when states are localized, but it relies in an essential way on localization when considering larger
and larger systems, where the variational Koopmans' orbitals converge rapidly to their thermodynamic limit (for the sake of illustration, these can be thought to closely resemble maximally localized Wannier functions~\cite{RevModPhys.84.1419}). This point will be discussed in detail in the second part of the paper.

The fact that Koopmans' compliance can lead to orbital energies that can be compared to the quasiparticle excitation energies of photoemission experiments, and to
canonical orbitals that resemble Dyson orbitals, has been discussed extensively for the case of molecular systems~\cite{ferr+14prb}: In previous work~\cite{Dabo2013,nguyen_first-principles_2015,Nguyen_JCTC_2016} we presented the performance of KC functionals in predicting frontier energies, ultraviolet photoemission spectra, and orbital tomography momentum maps for
different classes of molecules, while also arguing that these functionals represent quasiparticle approximations to the exact spectral functional~\cite{ferr+14prb, Gatti2007prl}. In fact, we typically find very good agreement with experiments, comparable or sometimes even better than state-of-the-art MBPT methods, while preserving moderate computational costs and the quality of the potential energy surfaces of the underlying base functionals~\cite{Borghi_PRB_2014}, or even improving on these when the KIPZ functional is used, thanks to its local self-interaction removal on the total energy~\cite{Nguyen_JCTC_2016}. 

In this work, we discuss how the framework of KC functionals extends to the case of solids. We focus first on the conceptual issues, and then on the calculation of energy gaps and IP energies. In Section~\ref{sec:theory} we describe the main theoretical challenges and the approach adopted in this work. In Section~\ref{sec:results_1D} we study finite alkane chains of increasing length, and discuss the thermodynamic limit in these one-dimensional systems. Then, in Section~\ref{sec:results_3D} we assess the method against the calculation of band gaps in 3D semiconductors and insulators as well as selected surfaces. The accuracy in predicting  E$_{\rm g}$ and IP energies is compared to experiments, standard KS-DFT calculations, many-body perturbation theory, and coupled-cluster [CCSD(T)] wavefunction methods. 
 
\section{Theory and Methods}
\label{sec:theory}
%
\subsection{Linearization in Koopmans-compliant functionals}
%
As mentioned, KC functionals~\cite{dabo_towards_2008, Dabo2009, Dabo2010, psik_koopmans, dabo_piecewise_2014,Dabo2013,Borghi_PRB_2014} explicitly enforce PWL conditions to an entire electronic manifold, introducing energy functionals that
depend linearly on occupation numbers.  Formally, these functionals are constructed by removing, orbital-by-orbital, the non-linear (Slater) contribution to the total energy and by replacing it by a linear (Koopmans) term.
The slope of the linear term can be chosen in a number of ways, leading to different KC flavors. In this work we focus on the KI and KIPZ implementations, described in detail in Ref.~\onlinecite{Borghi_PRB_2014}: In KI the slope is chosen as the total energy difference of two adjacent electronic configurations with integer occupations, and in KIPZ the same is done on the Perdew-Zunger (PZ) self-interaction corrected functional~\cite{perdew_self-interaction_1981}. Given an approximate (``base'') DFT functional $E^{\rm app}$, its KC counterpart can be written as:
\begin{equation}\label{kc_gen}
  E^{\rm KC} =  E^{\rm app} + \sum_{i} \alpha_i \Pi^{\rm KC}_i ,
\end{equation}
where
\begin{align}\label{Eq:KC_KIPZ} 
     \Pi^{\rm KI}_i &=  E_{\rm Hxc} [\rho-\rho_i] -E_{\rm Hxc}[\rho] \nonumber \\      
     &+f_i \Big[ E_{\rm Hxc}[\rho-\rho_i+n_i] -E_{\rm Hxc}[\rho-\rho_i] \Big], \\
     \Pi^{\rm KIPZ}_i &= \Pi^{\rm KI}_i -f_i E_{\rm Hxc} [n_i],
\end{align}
having defined $\rho_i(\mathbf{r}) = f_i n_i(\mathbf{r}) = f_i |\varphi_i(\mathbf{r})|^2$, and $E_{\rm Hxc}$ being the Hartree and exchange-correlation energy corresponding to the underlying approximate DFT functional.
For all calculations presented in this work, the base functional is PBE.~\cite{perdew_generalized_1996}
As mentioned, the orbital-dependent factors $\alpha_{i}$ account for electronic screening and orbitals relaxation; for $\alpha_{i}=1$, the KC functional in Eq.~(\ref{kc_gen}) fulfills exactly the generalized Koopmans' theorem at frozen orbitals~\cite{psik_koopmans}.
A Koopmans' orbital-by-orbital linearity condition is more stringent than the piecewise linearity condition satisfied by the exact KS-DFT ground-state energy as a function of fractional changes in the occupation of the HOMO, and in turn provides a more general orbital-density dependent (ODD) framework. In fact, at variance with DFT functionals but similarly to
 other ODD methodsm such as the PZ self-interaction correction~\cite{perdew_self-interaction_1981}, KC functionals are not invariant under unitary transformations within the manifold of filled orbitals.~\cite{hofmann_using_2012,lehtola_variational_2014,ferr+14prb,Borghi_PRB_2015}
The {\it variational orbitals} $\{\ket{\varphi_i}\}$ that minimize the functional are therefore different from the eigenstates or {\it canonical orbitals} $\{\ket{\phi_m}\}$ that diagonalize the KC Hamiltonian, as discussed e.g. in Refs.~\onlinecite{hofmann_using_2012,lehtola_variational_2014,Borghi_PRB_2015}.
The algorithm that we advocate to minimize the KC functional consists of two nested steps~\cite{Borghi_PRB_2015}, following the ensemble-DFT approach~\cite{marzari_ensemble_1997}:
($i$) a minimization with respect to unitary transformations at fixed orbital manifold (inner loop), that enforces the Pederson condition~\cite{pederson-jcp-1984} and leads to a projected, 
unitary-covariant functional of the orbitals, within
($ii$) a variational optimization of the orbital manifold of this
projected functional (outer loop). 

Koopmans' compliance from Eqs.~(\ref{kc_gen}-\ref{Eq:KC_KIPZ}) can be imposed to both valence and conduction states. Currently, the only requirement is that the system under consideration needs to have a finite gap, which ensures that the occupation of any variational orbital is either zero or one, and the definition of orbital densities $\rho_i(\mathbf{r})$ is unambiguous. In this case, any matrix elements between filled and empty state introduced by the Koopmans corrections are projected out and the minimization can be performed separately for each manifold. 
Imposing the Pederson condition to the occupied states leads to a set of well localized variational orbitals, when using the KIPZ functional, or the KI functional taken as a limit of the KIPZ with zero PZ correction~\cite{Borghi_PRB_2014}, while it
often tends to delocalize the empty states.~\cite{empty_delo}
Here, we stress that calculating the Koopmans' corrections on a localized set of orbitals $\{\ket{\varphi_i}\}$ is a key requirement to deal with extended systems. Indeed, in the limit of a crystalline system the canonical orbitals $\ket{\phi_m}$ satisfy Bloch theorem, while the variational orbitals $\{\ket{\varphi_i}\}$ corresponding to the occupied states remain localized (and are actually very similar to maximally localized Wannier functions~\cite{RevModPhys.84.1419} ). 
The $\Pi^{\rm KC}_i$ correction is therefore still finite and depends only on local properties, even in periodic crystals, and converges rapidly to its non-trivial thermodynamical limit. 

A workaround for the delocalization of empty states in the current functionals is to compute a non-self-consistent Koopmans-correction using maximally localized Wannier orbitals as the localized representation for the lower part of the manifold of empty states~\cite{mwlf_emp}. 
Even though this choice is arbitrary, it can provide a practical and effective scheme, as clearly supported by the results of the present work. 

\subsection{Screening in Koopmans-compliant functionals}
%
As mentioned above, the screening factors $\alpha_i$ in the definition of KC functionals account for orbitals relaxation and should be orbital-dependent, and applied to the variational orbitals that minimize the functional. They can be calculated by evaluating an orbital-dependent KS-specific linear-response function~\cite{psik_koopmans, Nicola_unpublished} employing the machinery of density-functional perturbation theory.~\cite{RevModPhys.73.515} A simpler approach can also be used, based on the linearization of the energy as a function of the single-particle occupation numbers; this is the approach that will be followed here.
We remind that for small molecules~\cite{nguyen_first-principles_2015,Nguyen_JCTC_2016} it is typically sufficient to compute two values of $\alpha$, one for occupied and the other for empty states, and these can be chosen by enforcing that the HOMO eigenvalue of a neutral molecule is equal to the LUMO eigenvalue of the respective cation, and that the HOMO eigenvalue of an anion molecule is equal to the LUMO eigenvalue of the neutral, respectively. However, the screening calculated on frontier (canonical) orbitals is only meant to act as an average measure of the response of the electronic system at hand; in an orbital-resolved framework it should be applied to the
individual variational orbitals (not to mention that in the solid state limit there is no difference e.g. in the HOMO for the neutral and the LUMO for the cation).

The finite-difference procedure we adopt here for the calculation of the screening parameter $\alpha_i$ corresponding to a given orbital $\varphi_i$ takes place in two steps, using an {\it auxiliary} system where the occupation $f_i$ of the orbital $\varphi_i$ is variable;
a linear-response approach would be more elegant, and its implementation is currently under way \cite{Nicola_unpublished}, but the protocol below can always be used when linear-response techniques are not available.

Without loss of generality, in the following discussion we assume that the orbital under consideration belongs to the occupied manifold of the {\it original} system. The goal is to determine each screening coefficient $\alpha_i$ such that the expectation value of the Koopmans' Hamiltonian on the variational orbital under consideration is independent on its own occupation:
\begin{equation}
\left.\frac{d E^{\rm KC}}{df_i}\right|_{f_i=\bar{f}} = \left.\langle\varphi_i |H^{\rm app}+ \alpha_i \hat{v}_{i} |\varphi_i\rangle \right|_{\bar{f}} = \lambda_{ii}^{\alpha_i} = \text{constant in } \bar{f},
\label{janak_lam}
\end{equation}
where the first identity is a generalization of the Janak theorem~\cite{PhysRevB.18.7165} for ODD functionals, and $\hat{v}_{i}(\bf r)=\frac{\delta \Pi^{\rm KC}(\rho_{i})}{\delta \rho_{i}(\bf r)}$ is the ODD potential. 
In practice, this is achieved trough the procedure illustrated below.

{\it Step 1.} For a given value of $f_i=\bar{f}\in [0,1]$, we minimize the KC total energy, starting with a best-guess trial value of $\alpha = \alpha^{(0)}$ identical for all states. In order to avoid that $\varphi_i$ morphs into the VBM (this would always be the most favorable solution because of the Aufbau principle), it is kept frozen during the minimization while imposing the standard orthogonality condition with all other states belonging to the same spin channel (a similar treatment has been recently introduced by Ma and Wang~\cite{ma_using_2016} for computing band gaps using projected Wannier orbitals). For states in the opposite spin channel a standard optimization of the orbitals is performed. At the end of the optimization the minimum total energy $E^{\rm KC}_i(\bar{f})$ compatible with the constraints imposed is obtained and the expectation value $\lambda_{ii}^{\alpha^{(0)}}(\bar{f})$ of the KC Hamiltonian on $\varphi_i$ is calculated.
Typically, we repeat this constrained minimization for two values $\bar{f}=0$ and $\bar{f}=1$, with the wavefunctions of the {\it auxiliary} system initialized to those of the {\it original} system computed with the trial screening factor $\alpha^{(0)}$.

{\it Step 2.} We search for the optimal value of $\alpha_i$ for which $\lambda_{ii}^{\alpha_i}(0) = \lambda_{ii}^{\alpha_i}(1)$. Within a second-order approximation of the total energy as a function of $f_i$, and assuming a linear dependence of $\lambda_{ii}$ on $\alpha_i$, this condition leads to the following expression for $\alpha_i$:
\begin{equation}
   \alpha_i = \alpha^{(0)} \frac{\Delta E_i - \langle\varphi_i |H^{\rm app}|\varphi_i\rangle|_0}{\lambda^{\alpha^{(0)}}_{ii}(0) - \langle\varphi_i |H^{\rm app}|\varphi_i\rangle|_0},
\end{equation}
where $\Delta E_i = E_i^{\rm KC}(\bar{f}=1) - E_i^{\rm KC}(\bar{f}=0)$.
This two-step procedure is applied to compute $\alpha_i$ for each variational orbital~\cite{screening_note}
In practice, the number of calculations for $\alpha_i$ can be greatly reduced by exploiting the symmetry of the variational orbitals - e.g.
in bulk silicon there is only one kind of variational orbital, similar to a bonding Wannier function. 
Moreover, because each $\alpha_i$ can be computed independently, these calculations can be run trivially in parallel.

It is relevant to note that due to the finite size of the supercells considered, the variation of $\bar{f}$ in  Step 1 produces a spurious interaction between an additional charge density and its periodic replicas, lowering both $E^{\rm KC}_i$ and $\lambda_{ii}^{\alpha_i}$. In order to correct for this we used three-dimensional (3D) real-space counter-charge corrections~\cite{dabo_electrostatics_2008,li-dabo11prb} for the study of isolated molecules, an image-charge interaction correction model based on the generalized Makov-Payne method,~\cite{makov-payne1995} in case of 3D crystals, and extrapolations in one dimension and counter-charges in the other two dimensions for the infinite polyethylene chain (detailed expressions for these corrections are provided in SM~\cite{Supplemental_material}).

\subsection{Localization in Koopmans-compliant functionals, and the thermodynamic limit}
\label{sec:results_1D}
\begin{figure*} 
  \includegraphics[width=0.9\textwidth]{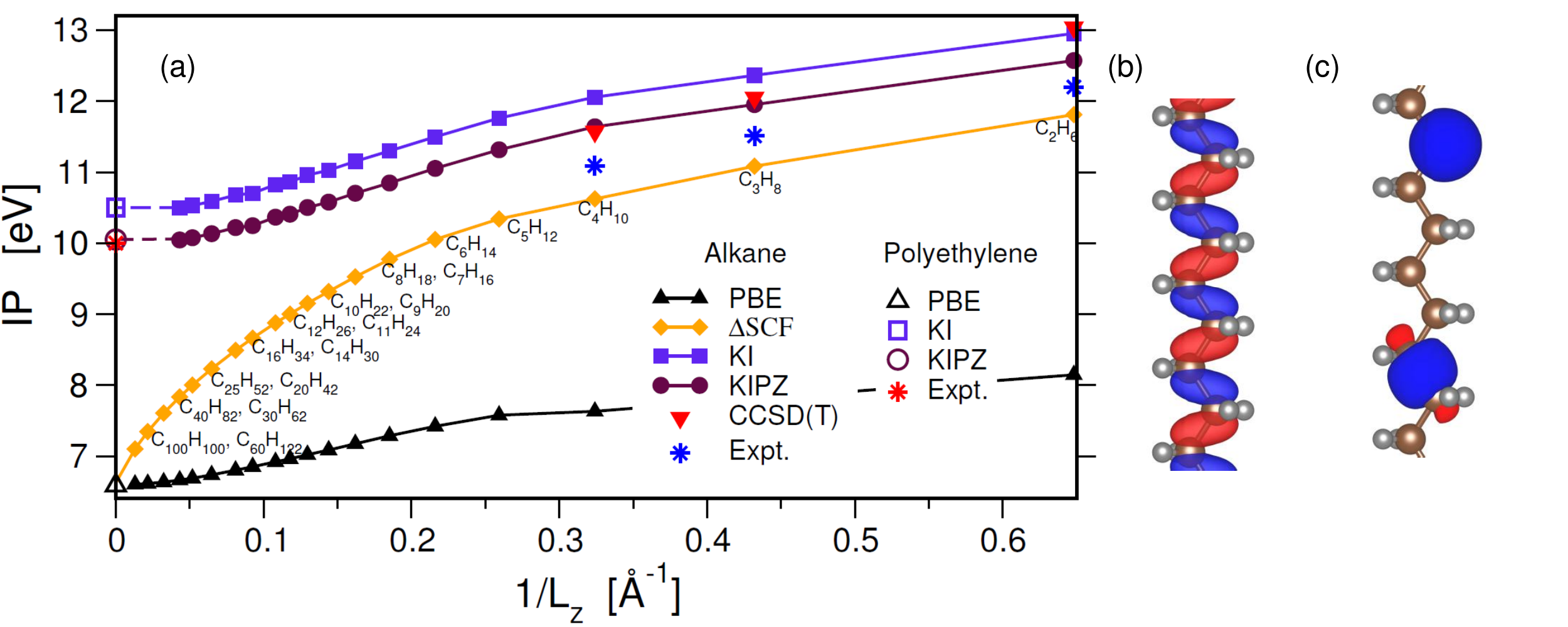}
  \caption{\label{fig_ip_alkanes} 
  Panel (a): IP energy as opposite of the HOMO for oligo-alkanes as a function of the inverse of the system length (L$_z$ in \AA), and computed with DFT-PBE, $\Delta$SCF, KI, and KIPZ, together with polyethylene (infinite chain). For $\Delta$SCF the results at L$_z \rightarrow \infty$ is an extrapolation.
  Right panels show (b) the VBM (canonical) orbital of polyethylene computed with DFT-PBE and (c) two types of variational orbitals for the same system computed using KIPZ, and very similar maximally localized Wannier functions for C-C and C-H $\sigma$ bonds.
  }
\end{figure*}

In order to investigate how KC functionals work on extended systems we start 
by calculating the IPs, as the opposite of the HOMOs, in linear alkane molecules (C$_n$H$_{2n+1}$--where $n > 1$) of increasing length. In particular, we study how the IPs change as a function of molecular length towards the thermodynamic limit represented by the infinite polyethylene (PE) chain. 
This example also clarifies the importance of localization when working with KC functionals.

Straight-alkane chains with staggered conformation are studied in orthorhombic supercells with at least 15~\AA~of separation in each direction. 
To study the infinite polyethylene chain, we consider a supercell containing a continuous C$_{n}$H$_{2n}$ 1D chain aligned along $z$. 
A value of $n=22$ (corresponding to a periodic supercell with the long side of $\simeq 28.5$~\AA) and $\Gamma-$point sampling for Brillouin zone integration has been used. This is equivalent to a $1\times1\times11$ $\mathbf{k}$-point mesh for the primitive cell with formula unit C$_2$H$_4$. 

Details on the convergence of the IP as a function of vacuum size along these directions, as well as on the correction for the finite-cell effects in the $z-$direction when computing the screening factors $\alpha_i$ are discussed in SM~\cite{Supplemental_material}. Calculations have been done in a plane-wave basis set using norm-conserving pseudopotentials~\cite{Schlipf201536} to describe ion-electron interactions. The kinetic energy cutoff for wave functions has been set to 80 Ry. 

We show in Fig.~\ref{fig_ip_alkanes} the IP energies for 19 alkanes ($n$ going from 2 to 100) and for the extended polyethylene chain, computed by using DFT-PBE, KI, and KIPZ. We remind here that the IP energy is defined as the negative of the KS-HOMO (or VBM) energy with respect to the vacuum level. For all the finite molecules, we also provide the $\Delta$SCF results, where the IP is computed as the total energy difference between the neutral molecule and its cation, both at the PBE level. Experimental and CCSD(T) results are also shown (these latter are only available for the three smallest alkanes).~\cite{doi:10.1080/00268976.2015.1025113}

We find that the KI and KIPZ mean absolute errors (MAE) with respect to experiments are about 0.86 and 0.45 eV, respectively. This accuracy is comparable with that of CCSD(T), that has a MAE of about 0.62 eV. In contrast, DFT-PBE significantly underestimates the IP for these molecules with a MAE about of 3.74 eV, reflecting the intrinsic self-interaction error present in this functional. 

As expected, the performance of $\Delta$SCF shows a strong dependence on system size. For the three smallest molecules, $\Delta$SCF predicts IPs with an accuracy equivalent to that of KC functionals (i.e., MAE $= 0.43$ eV). However, increasing the size of the alkanes, the discrepancies become more and more significant, and, as is well known, when the size approaches the thermodynamic limit the $\Delta$SCF IP reduces to the PBE one~\cite{godby_density-relaxation_1998,sham_density-functional_1983,perdew_physical_1983}.
The failure of $\Delta$SCF in the thermodynamic limit has been discussed extensively e.g. for silicon nanocrystals (see Ref.~\onlinecite{godby_density-relaxation_1998} and associated discussion) and for hydrogen chains~\cite{Mori-Sanchez_2008}; some of the subtler reasons related to its application to the exact or approximate (e.g. local) functionals are still debated.~\cite{PhysRevB.91.245120,Mori-Sanchez_2008,Kronik_JCP_2015}.
In a nutshell, in an approximate functional and as the length of the molecule increases, the HOMO orbital becomes delocalized along the chain [see Fig.~\ref{fig_ip_alkanes}(a)]. Removing an electron from this orbital --- which is exactly what happens in a $\Delta$SCF calculation --- only slightly modifies the local value of the charge density since the orbital is normalized to one when integrating over the entire system. 
In the limit of an infinite system the $\Delta$SCF IP then reduces to the derivative of the total energy with respect to the particle number ~\cite{Mori-Sanchez_2008}, which, for a local or semilocal density functional approximation, is the negative of the KS-DFT HOMO eigenvalue.~\cite{cohen_fractional_2008} (note that for the exact functional, the IP would be correct, as would be the band gap calculated as $E_{N+1}+E_{N-1}-2E_N$~\cite{godby_density-relaxation_1998,sham_density-functional_1983}).

Two possible routes to overcome these limitations are e.g. going beyond the local or semi-local nature of the approximate functionals (e.g. having a non-local second derivative with respect to the density~\cite{Kronik_JCP_2014, PhysRevB.94.075123} or modeling the discontinuity of the KS potential~\cite{kuisma_kohn-sham_2010}), or retaining the simplicity of standard density-functional approximations and working in a localized representation of the orbitals. The dielectric screening localization suggested by Chan and Ceder~\cite{chan_efficient_2010}, leading indeed to satisfactory prediction of fundamental band gaps in solids, is an example of the feasibility of this second route.
Since in KC functionals the generalized Koopmans' condition is imposed on the {\it variational} orbitals, which are localized, rather than the {\it canonical} ones [as shown in Fig.~\ref{fig_ip_alkanes}(b)], a non-zero correction is present also in the thermodynamic limit. In fact, KI and KIPZ calculations predict  the IP energy of polyethylene to be 10.50 and 10.07 eV, respectively, in very good agreement with early experimental estimates~\cite{Partridge_JCP_1966} (about 10.0 eV); PBE underestimates these values by about $3.5$ eV.
 
\section{Results and Discussion}
\label{sec:results}
%
\subsection{Band energies for solids}
\label{sec:results_3D}
\begin{table}
  \caption{\label{error_gap_ip} Mean absolute errors (MAE, in eV) and mean absolute percent errors (MAPEs, in \%) with respect to experiments for: ($i$) band gaps of 16 solids presented in Fig.\ref{fig_gaps} for which experimental, G$_0$W$_0$ and the quasiparticle self-consistent GW with vertex-correction (QSG\~{W}) data are available 
  (see Table II of S.M.~\cite{Supplemental_material}, and Refs.~\onlinecite{PhysRevB.75.235102, PhysRevLett.99.246403, PhysRevB.92.041115}), and ($ii$) IP of 6 surfaces presented in Fig.~\ref{fig_ip_6surf}. Experimental values for solid band gap and surfaces are taken from Refs.~\onlinecite{Madelung_book, PhysRevB.13.5530, PhysRevLett.34.528, PhysRevB.92.041115}, respectively.}
  \begin{ruledtabular}
  \begin{tabular}{l l r r r r r  }
              &           & PBE    &G$_0$W$_0$ & KI & KIPZ & QSG\~{W} \\
  \hline 
  $E_{\rm g}$ & MAE (eV)  &  2.54  &  0.56  & 0.27 & 0.22  & 0.18 \\ 
   	          & MAPE (\%) & 48.28  & 12.10  & 7.09 & 5.37  & 4.46 \\
  ${\rm IP}$  & MAE (eV)  &  1.09  &  0.39  & 0.19 & 0.21  & 0.49 \\ 
	          & MAPE (\%) & 15.58  & 5.71   & 2.99 & 3.14  & 7.41 \\ 
  \end{tabular}
  \end{ruledtabular}
\end{table}
\begin{figure}
  \includegraphics[width=0.45\textwidth]{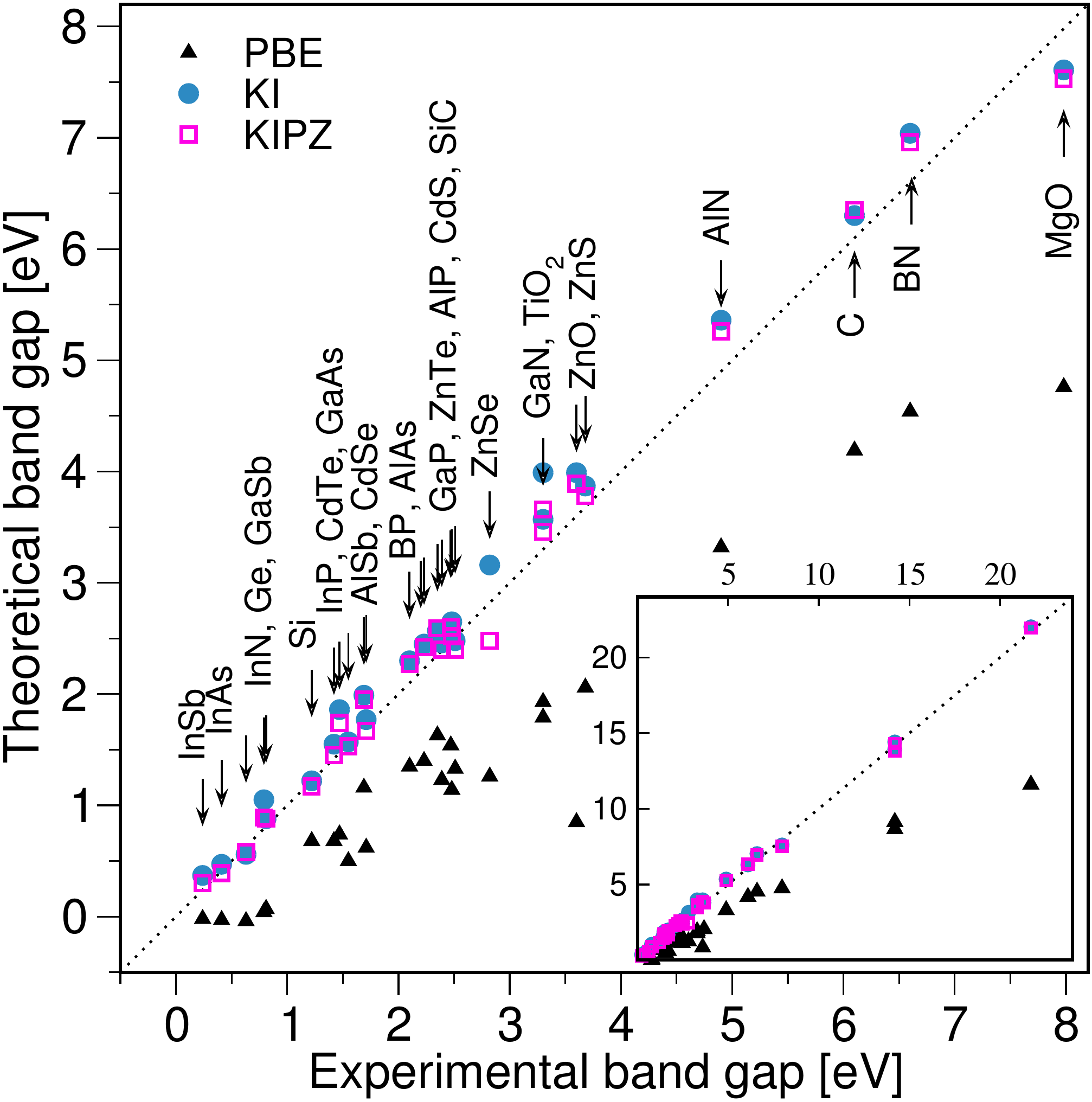}
  \caption{\label{fig_gaps} Band gaps of 30 semiconductors and insulators, calculated {using PBE, KI, or KIPZ functionals}, compared with available experimental data, shown in an energy range between -0.5 and 8.2 eV. A wider energy range (up to 22 eV) is shown in the inset.}
\end{figure}
\begin{figure} 
  \includegraphics[width=0.45\textwidth]{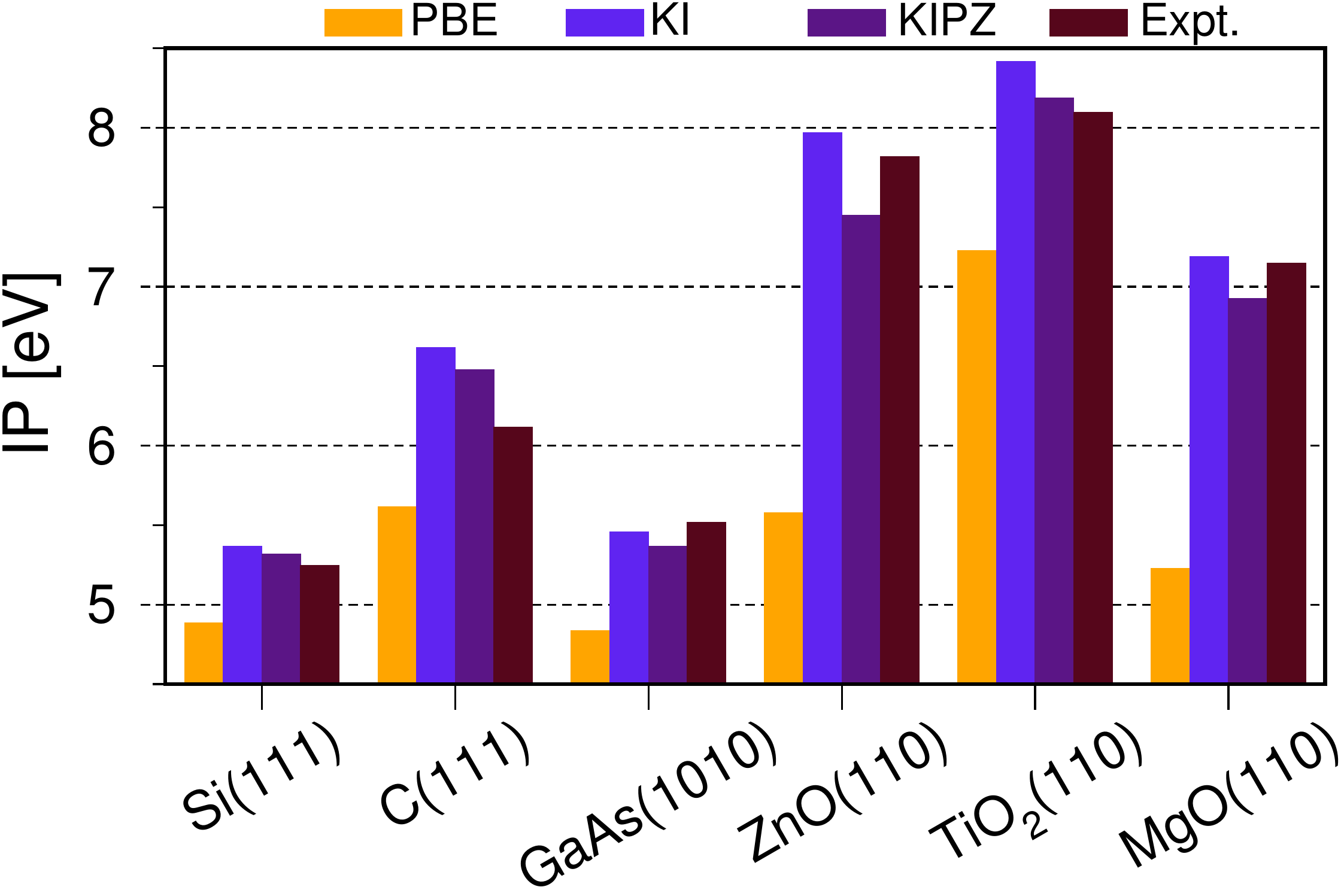}
  \caption{\label{fig_ip_6surf} Ionization potential (in eV) of six different surfaces, including Si(111), C(111), GaAs(1010), ZnO(10$\bar{1}$0), TiO$_2$(110) and MgO(110), calculated by using PBE, KI, and KIPZ. The results are compared to experimental values taken from Ref.~\onlinecite{PhysRevB.92.041115}.}
\end{figure}
Having described how KC functionals work in extended systems, we apply such formalism to predict the band gap E$_{\rm g}$ for a set of 30 compounds, including small gap semiconductors and large gap insulators for which accurate experimental and theoretical reference results are available. In Fig.~\ref{fig_gaps} and Table~\ref{error_gap_ip} we show the accuracy of the KC methods as compared with experiments, standard DFT-PBE , and state-of-the-art many-body perturbation theory methods.
The zero-point contribution is removed from the experimental data, when available (see SM~\cite{Supplemental_material} for the complete list).
The results show that in PBE the value of the E$_{\rm g}$ is underestimated with a MAE and mean absolute percent error (MAPE) of about 2.5 eV and 50\% with respect to experiments, respectively. Notably, MAE in KI and KIPZ are down to 0.27 and 0.22 eV; this latter is 
comparable with that obtained by quasi-particle self-consistent GW with vertex-corrections [using an effective exchange-correlation kernel $f_{xc}$ obtained from the Bethe-Salpeter equation (BSE)~\cite{PhysRevLett.99.246403}, or the bootstrap approximation,~\cite{PhysRevB.92.041115} to account for vertex-corrections], and more than twice as accurate as G$_0$W$_0$[PBE]~\cite{PhysRevB.92.041115,PhysRevB.75.235102} with a MAE of 0.56 eV.
In the present calculations the experimental lattice constants are used for consistency; also a supercell technique where the Brillouin zone integration is sampled only at the $\Gamma$ point is used. The band gap convergence with respect to supercell size has been tested, yielding an error bar smaller than 0.06 eV. Detailed convergence studies with respect to supercell size and cutoff energy for the plane wave expansion of wave functions for each system 
are presented in SM.~\cite{Supplemental_material}

To further stress how the KC functionals perform to correct the band gap of extended systems, we consider ZnO as a paradigmatic case study: this system is known to be a difficult case to deal with at the GW level, requiring a large number of empty states and dense $\mathbf{k}$-point sampling for the calculation of quasiparticle corrections.~\cite{PhysRevLett.105.146401,PhysRevB.84.241201}
For this system, the use of KC functionals has two main effects: first, $d$ states are shifted to a more accurate position with respect to the top of the valence band, and second, the fundamental gap is enlarged with respect to that of PBE, in much better agreement with experiments. 
The KI and KIPZ predictions of E$_g$ for ZnO are about 3.96 and 3.76 eV, respectively, close to the experimental value of 3.6 eV when the zero-point renormalization correction is considered (3.44 eV otherwise). 
During the KC minimization, three classes of variational orbitals (corresponding also to three different  $\alpha_i$ screening values) have been found, namely
the $d$-states centered at Zn atoms, the $\sigma$-like states localized near O atoms and the $\sigma^*$-like states near Zn.
From the analysis of the projected density of state, we find that with the optimal $\alpha_i$ values the energy levels of the $d-$band center computed by KI and KIPZ are about 6.81 eV and 7.00 eV with respect to the VBM, which are close to the experimental values($7.50-8.81$ eV) and to that predicted by DFT+$U$, 9.00 eV~\cite{PhysRevX.5.011006}, and larger than the PBE prediction of 5.10 eV. KI and KIPZ results. Here, one should also note that the DFT+$U$ prediction for the $d-$band center strongly depends on the chosen Hubbard $U$ parameters (see Ref.~\onlinecite{PhysRevX.5.011006} and references therein).  

Overall, we find that KIPZ performs sligthly better than KI in predicting E$_{\rm g}$. This can be explained by the fact that the KIPZ functional is able to modify not only the electronic excitation energies of approximate DFT functionals, but also the manifold of electronic orbitals (i.e. the single-particle KS density-matrix).
In particular, a more accurate spatial decay of the density matrix is usually expected as a result of imposing PWL via KIPZ~\cite{Borghi_PRB_2014}.
It is relevant to note that a side-effect of having a finite variational PZ term in KIPZ, or infintesimal one in KI, leads to a small symmetry breaking and splitting of the $d$ levels; this unphysical broken symmetry is driven by the  lack of rotational invariance of the PZ functional. In this case it one should note that it does not even affect the band gap since this is derived from the $s-$ and $p-$like orbitals~\cite{Pederson_JCP_2014}).

\subsection{Surfaces: Determination of band edges}
\label{sec:surfaces}

Besides the fundamental gap, the accurate determination of band edge positions is also very important, affecting, e.g. the band alignment at interfaces. In practice, band edge positions can not be extracted directly from periodic bulk calculations, since an absolute energy reference is actually needed. 
A viable solution is to make reference to the IP calculated through the use of surface slabs. The IP is formally defined as the energy difference between the vacuum level $E_{\rm vac}$ and the valence band maximum (VBM). To simplify the convergence of results with respect to the slab thickness, the VBM is determined in a separate bulk calculation and referenced to a local reference potential, $V^{\rm b}_{\rm ref}$, which corresponds to the sum of the electrostatic potential and the local pseudopotential term~\cite{baldereschi_band_1988,singh-miller_surface_2009}. The IP is therefore calculated as
\begin{equation}\label{ip_form}
 {\rm IP} = (E_{\rm vac} - V^{\rm s}_{\rm ref}) - (\epsilon^{\rm b}_{\rm VBM} - V^{\rm b}_{\rm ref})
\end{equation}
where the superscript ``s'' (``b'') refer to slab (bulk) calculations. 
The slab system is taken sufficiently thick so that the local reference potential inside the slab, $V^{\rm s}_{\rm ref}$, corresponds to $V^{\rm b}_{\rm ref}$. In practice, $E_{\rm vac}$ and $V^{\rm s}_{\rm ref}$ are commonly determined at the PBE level, which 
has been shown to be reasonably accurate in comparison with higher level methods such as hybrid functionals~\cite{PhysRevLett.101.106802,PhysRevB.89.205309} or GW.~\cite{PhysRevLett.100.186401}

Using the above definitions, we carried out calculations of IPs for six different surfaces, including Si(111), C(111), GaAs(1010), ZnO(10$\bar{1}$0), TiO$_2$(110), and MgO(110). For comparison, we have used the same surface geometries as in Ref.~[\onlinecite{PhysRevB.92.041115}]: 13 atomic layers for GaAs(1010), 12 atomic layers for the TiO$_2$ and MgO (110) surfaces, and 24 atomic layers for the (111) reconstructed surface of Si, Ge, and diamond. The thickness of the vacuum region (25 \AA) has been chosen to minimize the interactions between periodic images, and has been kept the same for all slabs. 
Results are reported in Fig.~\ref{fig_ip_6surf} and Table~\ref{error_gap_ip}, 
showing that the accuracy of the KC functionals is not only much better than PBE, but also compares favorably with the G$_0$W$_0$ and QSG\~{W} methods (performed on the same geometries~\cite{PhysRevB.92.041115}), being basically twice as accurate. Here, QSG\~{W} uses an approximate bootstrap exchange-correlation kernel~\cite{PhysRevB.92.041115} to account for vertex-corrections. 
It is noteworthy that the KI method does not change the ground-state density of the base functional, making KI $E_{\rm vac}$ and $V^{\rm s}_{\rm ref}$ identical to those computed at PBE level. 
This is not true for the KIPZ functional, and to overcome this inconsistency one might compute also $E_{\rm vac}$ and $V^{\rm s}_{\rm ref}$ at the KIPZ level; work in this direction could be considered for a future study.  

\section{Conclusions}

We have investigated the extension of Koopmans-compliant (KC) functionals to the case of extended systems,
comparing the results with experimental data and state-of-the art many-body perturbation theory for a broad range of well characterized semiconductors and insulators.
In doing so, we have developed a new approach to compute the orbital-dependent screening factors for KC functionals, and applied it to predict the IP and band gaps for both finite and extended systems.
First, we have discussed KC functionals for the case of one-dimensional systems  with increasing sizes and reaching the thermodynamic limit.
This analysis has reiterated the importance of imposing the criterion of piecewise linearity on localized orbitals, at variance with the $\Delta$SCF approach, which can only use canonical orbitals to predict IPs and EAs and breaks down in extended systems.
Then we have studied reference solids and shown that the KI and KIPZ functionals can yield very accurate results for band gaps of different semiconductors and insulators, with mean absolute errors that are of the order of 0.2 eV and comparable with the most accurate SQG\~{W} data available. The comparison is even more favorable for the IPs studied with an accuracy doubled with respect to SQG\~{W}. These results are ever more remarkable considering that the physics of the problem remains that of the PBE generalized-gradient approximations,
and have been obtained with a functional theory of the occupied states. This accuracy and simplicity, given the computational costs broadly comparable to standard density-functional theory, makes KC functionals very attractive to study electronic levels in complex materials and devices (provided a band-gap is maintained). It also  reiterates the suggestion that charged excitations, such as electron additions and removals, can be studied with functional theories that are dynamical~\cite{ferr+14prb,Gatti2007prl} (i.e. frequency-dependent); thus, KC functionals take the role of spectral functionals, and offer a quasiparticle approximation to the exact one~\cite{ferr+14prb}.

\begin{acknowledgments}
We acknowledge partial support from the Swiss National Centre
for Computational Design and Discovery of Novel Materials (MARVEL) and from the EU Centre of Excellence ``MaX - Materials Design at the Exascale'' (Horizon 2020 EINFRA-5, Grant No. 676598).
We would like to thank Prof. Ismaila Dabo, Dr. Marco Gibertini, and Dr. Fabien Bruneval for useful discussions, and Dr. Wei Chen and Prof. Alfredo Pasquarello for providing the surface geometries used in Sec.~\ref{sec:surfaces}.
\end{acknowledgments}

%
\bibliography{biblio}

\begin{thebibliography}{10}
\bibitem{mwlf_emp}
To obtain the maximally localized Wannier orbitals for the low-lying empty
  states, we employed the disentanglement maximally localized Wannier function
  technique proposed in Ref.~\onlinecite{PhysRevB.65.035109}.
\bibitem{screening_note}
For empty states the procedure is essentially the same. The only difference is
  that in this case we have an extra (partially) occupied state in the
  auxiliary system with $N+f_i$ electrons corresponding to the empty state in
  the original N-electron system.
\end{thebibliography}
\bibliographystyle{aip}
\end{document}